\documentclass[twocolumn]{aastex631}

\newcommand\N[1]{{\color{blue} \bf #1}} %N8
\newcommand\J[1]{{\color{purple} \bf #1}} %N8
\usepackage{soul}

\shorttitle{The Remains of SN\,2015bh}
\shortauthors{Jencson et al.}
\graphicspath{{./}{figures/}}
\begin{document}

\title{Hubble Space Telescope Imaging Reveals That SN\,2015bh Is Much Fainter than Its Progenitor}

\correspondingauthor{Jacob E.\ Jencson}
\email{jjencson@email.arizona.edu}

\author[0000-0001-5754-4007]{Jacob E. Jencson}
\affil{Steward Observatory, University of Arizona, 933 North Cherry Avenue, Tucson, AZ 85721-0065, USA}

\author[0000-0003-4102-380X]{David J. Sand}
\affil{Steward Observatory, University of Arizona, 933 North Cherry Avenue, Tucson, AZ 85721-0065, USA}

\author[0000-0003-0123-0062]{Jennifer E. Andrews}
\affil{Gemini Observatory/NSF's NOIRLab, 670 N. A'ohoku Place, Hilo, Hawai'i, 96720, USA}

\author[0000-0001-5510-2424]{Nathan Smith}
\affil{Steward Observatory, University of Arizona, 933 North Cherry Avenue, Tucson, AZ 85721-0065, USA}

\author[0000-0002-1468-9668]{Jay Strader}
\affil{Center for Data Intensive and Time Domain Astronomy, Department of Physics and Astronomy, Michigan State University, East Lansing, MI 48824, USA}

\author[0000-0001-8341-3940]{Mojgan Aghakhanloo}
\affil{Steward Observatory, University of Arizona, 933 North Cherry Avenue, Tucson, AZ 85721-0065, USA}

\author[0000-0002-0744-0047]{Jeniveve Pearson}
\affil{Steward Observatory, University of Arizona, 933 North Cherry Avenue, Tucson, AZ 85721-0065, USA}

\author[0000-0001-8818-0795]{Stefano Valenti}
\affil{Department of Physics and Astronomy, University of California, 1 Shields Avenue, Davis, CA 95616-5270, USA}



\begin{abstract}
We present Hubble Space Telescope (HST) imaging of the site of SN\,2015bh in the nearby spiral galaxy NGC\,2770 taken between 2017 and 2019, nearly four years after the peak of the explosion. In 2017--2018, the transient fades steadily in optical filters before declining more slowly to $F814W = -7.1$~mag in 2019, $\approx$4~mag below the level of its eruptive luminous blue variable (LBV) progenitor observed with HST in 2008--2009. The source fades at a constant color of $F555W - F814W = 0.4$~mag until 2018, similar to SN\,2009ip and consistent with a spectrum dominated by interaction of the ejecta with circumstellar material (CSM). A deep optical spectrum obtained in 2021 lacks signatures of ongoing interaction ($L_{\mathrm{H}\alpha} \lesssim 10^{38}$~erg~s$^{-1}$ for broadened emission $\lesssim$2000~km~s$^{-1}$), but indicates the presence of a nearby \ion{H}{2} region ($\lesssim$300\,pc). %Precise image alignments confirm that the position of the faint, unresolved $F814W$ source in 2019 is consistent with that of the transient.
The color evolution of the fading source makes it unlikely that emission from a scattered-light echo or binary OB companion of the progenitor contributes significantly to the flattening of the late-time light curve. The remaining emission in 2019 may plausibly be attributed an evolved/inflated companion or an unresolved ($\lesssim$3~pc), young stellar cluster. %\edit1{
Importantly, the color evolution of SN\,2015bh rules out scenarios in which the surviving progenitor is obscured by nascent dust and does not clearly indicate a transition to a hotter, optically faint state. %} 
The simplest explanation is that the massive progenitor did not survive. SN\,2015bh likely represents a remarkable example of the terminal explosion of a %\N{(N: very?  if progenitor was erupting, it might be only 20-30 Msun...you define "very massive" later as $>$70 Msun)} 
massive star preceded by decades of end-stage eruptive variability. 
\end{abstract}

\keywords{Supernovae(1375) --- Luminous blue variable stars(944) --- Massive stars(732) --- Stellar mass loss(1613) --- Evolved stars(481)}


%\section{List of Figures}
%\begin{enumerate}
%    \item Collage of HST images
%        \begin{enumerate}
%            \item F555W, F814W sequences
%        \end{enumerate}
%    \item Positional associations
%        \begin{enumerate}
%            \item F555W, F814W dolphot offsets.
%        \end{enumerate}
%    \item Light curves
%        \begin{enumerate}
%            \item HST, pre+post
%            \item ground-based (main-outburst)
%        \end{enumerate}
%    \item Spectrum (or lack thereof?)
%         \begin{enumerate}
%           \item Compare to progenitor or 09ip at similar phase?
%        \end{enumerate}
%   \item CMD (or HRD)
%          \begin{enumerate}
%            \item $F555W - F814W$ (or $V-I$)
%            \item Compare to other objects (09ip, LBVs?, etc.)
%        \end{enumerate}  
%\end{enumerate}

\section{Introduction} \label{sec:intro}
Many massive stars ($>$8~$M_{\odot}$) end their lives as core-collapse (CC) supernovae (SNe). Some fraction of massive stars may instead collapse directly to black holes, but the ranges of initial mass and evolutionary details that lead to these fates are debated. There is also mounting evidence that these cataclysms can be preceded by months to years of tumult. Type IIn SNe (the ``n'' signifies narrow emission features; \citealp{schlegel90,filippenko97,smith17_handbook}), in particular, show spectral signatures and high luminosities that require strong shock interaction with large masses of circumstellar material (CSM). This material may have been shed in violent eruptive events decades prior to the SN (see \citealt{smith14} for a review) %\edit1{
or in an enhanced wind in the last $\sim$10$^{3}$~yr of the progenitor's life \citep[e.g.][]{yoon10,smith17b}. %}
In some cases, as we discuss further below, outbursts have in fact been detected in the years directly preceding an SN. Massive star models connect this pre-SN variability to the late nuclear-burning phases that occur in the final few years of a massive star's life \citep{quataert12,smith14c,fuller17}. For very massive stars ($\gtrsim$70~$M_{\odot}$), the pulsational pair-instability mechanism predicts luminous and diverse nonterminal events \citep{woosley17}. 

Observationally, there is a diverse class of intermediate-luminosity transients found in nearby galaxies dubbed ``SN impostors,'' so named because they were originally seen in surveys for SNe.  Their spectra mimic Type~IIn SNe, although SN impostors typically have even narrower emission lines and lower luminosities \citep{smith11}. They have been interpreted as nonterminal events linked to luminous blue variable (LBV) stars \citep[e.g.,][]{vandyk00}, but the physical mechanisms behind these outbursts remain unsatisfactorily explained. 
An important test to distinguish true CC SNe from impostors is to obtain deep, late-time imaging to search for a surviving star.
%} %\N{(N: not to be annoying, but "ultimate arbiter" might be overdoing it.  if you got deep imaging in 1870, you would have concluded that Eta Car was a definitely a supernova... maybe say "important test" or something...)} JJ: fixed. 
This can be difficult to constrain in practice for extragalactic sources, as ejected material may form obscuring dust, while ongoing CSM interaction may resemble a surviving star or mask the usual signatures of a CC SN (see, e.g., the recent debate surrounding SN\,1961V; \citealp[][]{kochanek11a,smith11,vandyk12,patton19,woosley22}). 

Precursor outbursts during the months to years before an apparent SN explosion have now also been seen in several instances \citep[e.g.,][]{foley07,pastorello07,fraser13b,mauerhan13,ofek13,margutti14,ofek14,elias-rosa16,ofek16,tartaglia16,thone17,nyholm17,pastorello18,reguitti19,ho19,strotjohann21}. %\N{(N: I would either cite a review, or cite them all since there aren't that many.  just these 3 seems lacking.  also cite 15bh, 09ip, 61V, Gaia126cfr by Kilpatrick, and maybe that Ofek paper, although maybe the Strotjohann paper is enough for that)} 
Perhaps the most well-studied example of this phenomenon is the enigmatic SN\,2009ip. The object was first identified as a bright outburst in 2009 of a massive ($\gtrsim$50--60~$M_{\odot}$) blue supergiant (BSG) star \citep{smith10,foley11}. Subsequent observations revealed a series of outbursts and erratic variability that culminated in a much more luminous, Type IIn SN-like event in 2012 with broad 
%\N{(13,000 km/s according to mauerhan, at least in 2012... this can be confusing.  pre-SN outbursts had widths of 600 but with some fast 8,000 km/s stuff in absorption, but then 2012 SN event has broad 13,000 km/s P Cygni lines from SN ejecta...)} 
spectral features (FWHM $\approx$ 8000~km~s$^{-1}$, $\approx$13,000~km~s$^{-1}$ P Cygni absorption; \citealp{fraser13a,mauerhan13,pastorello13,prieto13,smith14b}). The main brightening of SN\,2009ip had a distinctive double-peaked light curve, the first reaching $M_R \approx -15$~mag and the second reaching $M_R \approx -18$~mag around 40\,days later. Monitoring of the event revealed the presence of dense, complex CSM \citep{graham14,margutti14,mauerhan14,martin15}, possibly with a disk-like geometry pointing to the influence of binary interactions \citep[e.g.,][]{mauerhan13,mauerhan14,levesque14,smith14,reilly17}. %\N{(N: many others said asymmetric and disk-like too - Mauerhan et al. 2013/2014, Smith et al.2014, Reilly et al. 2017 and others - based on polarimetry and line profiles, whereas Levesque's evidence for disk-like was just line ratios that suggested a high density and not a valid argument actually, because those high densities are always seen anyway...)} 
A class of objects with similar properties has been identified in recent years, including SN\,2010mc \citep{ofek13,smith13}, LSQ13zm \citep{tartaglia16}, AT\,2016jbu (or Gaia16crf; \citealp{kilpatrick18a,brennan22a,brennan22b}), SN\,2016bdu \citep{pastorello18}, and the main subject of this work, SN\,2015bh \citep{elias-rosa16,ofek16,thone17}.

The nature of these SN\,2009ip-like objects is debated and a range of physical scenarios has been proposed. %, including both terminal CC explosions and luminous outbursts that progenitor stars ultimately survive. 
In one nonterminal scenario, an eruptive outburst powers the first peak and subsequent interactions with previously ejected material powers the main peak \citep[e.g.,][]{pastorello13,fraser15,moriya15}. Comparisons have also been drawn between the rapid pre-SN variability of SN\,2009ip and the periastron collisions of an eccentric binary seen preceding the Great Eruption of $\eta$ Car \citep[e.g.,][]{smith11,smith11b}, in which case the main event may have been powered by the final merger %\edit1{
\citep{smith18,hirai21}. %} 
A merge-burst scenario has therefore also been suggested for SN~2009ip-like objects \citep{kashi13,soker13,soker16}. 
%something about asymmetry/torus for binary inter
Terminal scenarios have been proposed involving a relatively faint
% i deleted weak, because these authors said it was a 1e51 erg SN explosion.  faint becvause of blue progenitor, but not a weak explosion...
CC SN from a BSG (first peak) followed by strong circumstellar interaction (second peak) \citep{mauerhan13,smith14b}, with noted similarities between the first peak of SN\,2015bh and SN\,1987A \citep{elias-rosa16}. Alternatively, the first peak may represent a last-gasp precursor outburst followed by the final CC SN \citep{ofek13,tartaglia16,thone17}. An inherent feature in all of these scenarios is long-lived CSM interaction that dominates observations at late times, masking the ultimate fate of the progenitor. %\edit1{
Despite this complication, both SN\,2009ip and AT\,2016jbu have continued to fade below the level of their progenitors, indicating that the massive LBV stars are now gone \citep{smith22,brennan22c}.%}

Among the SN\,2009ip-like events, SN\,2015bh is remarkable for its exceptionally well-characterized progenitor and comprehensive monitoring of the evolving transient. The transient occurred in the nearby spiral galaxy NGC\,2770 (we adopt $D=28.8$\,Mpc; $m-M=32.3$\,mag, see Appendix~\ref{appendix:host}) with ample archival Hubble Space Telescope (HST) and ground-based data spanning decades. The pre-SN light curve of SN\,2015bh reveals a highly variable source since at least 1994 and episodes of rapid variability that were well documented in 2008, 2009, and 2013 \citep{elias-rosa16,ofek16,thone17,boian18}. A bright counterpart was identified in multiepoch, multifilter HST imaging in 2008--2009, indicating large variations in both temperature ($T \approx 5000$--$9000$\,K) and luminosity ($\log[L/L_{\odot}] \approx 5.9$--6.6) consistent with a very massive star experiencing LBV-like outbursts \citep{elias-rosa16}. A rare progenitor spectrum was obtained at the onset of the 2013 outburst that showed narrow H$\alpha$ emission ($\mathrm{FWHM} \lesssim 500$~km~s$^{-1}$) with a $\approx$1300~km~s$^{-1}$ P~Cygni absorption feature \citep{ofek16}. %\N{(N: but how broad were the emission lines?  That's where most of the mass is - and relevant to later narrow lines and CSM interaction...)} 
Based on radiative-transfer modeling of the spectrum, \citet{boian18} propose that the progenitor was an LBV, possibly $\gtrsim$35~$M_{\odot}$, with an optically thick wind.
%\N{(N: if it wasn;t exceeding the Eddington limit, which of course it might have been doing in eruption, so this isn't really a lower limit.)}. 
These prior studies note that if the event was nonterminal, the surviving star should be observable as a similarly luminous object in late-time imaging. 

\begin{figure*}
\centering
\includegraphics[width=\textwidth]{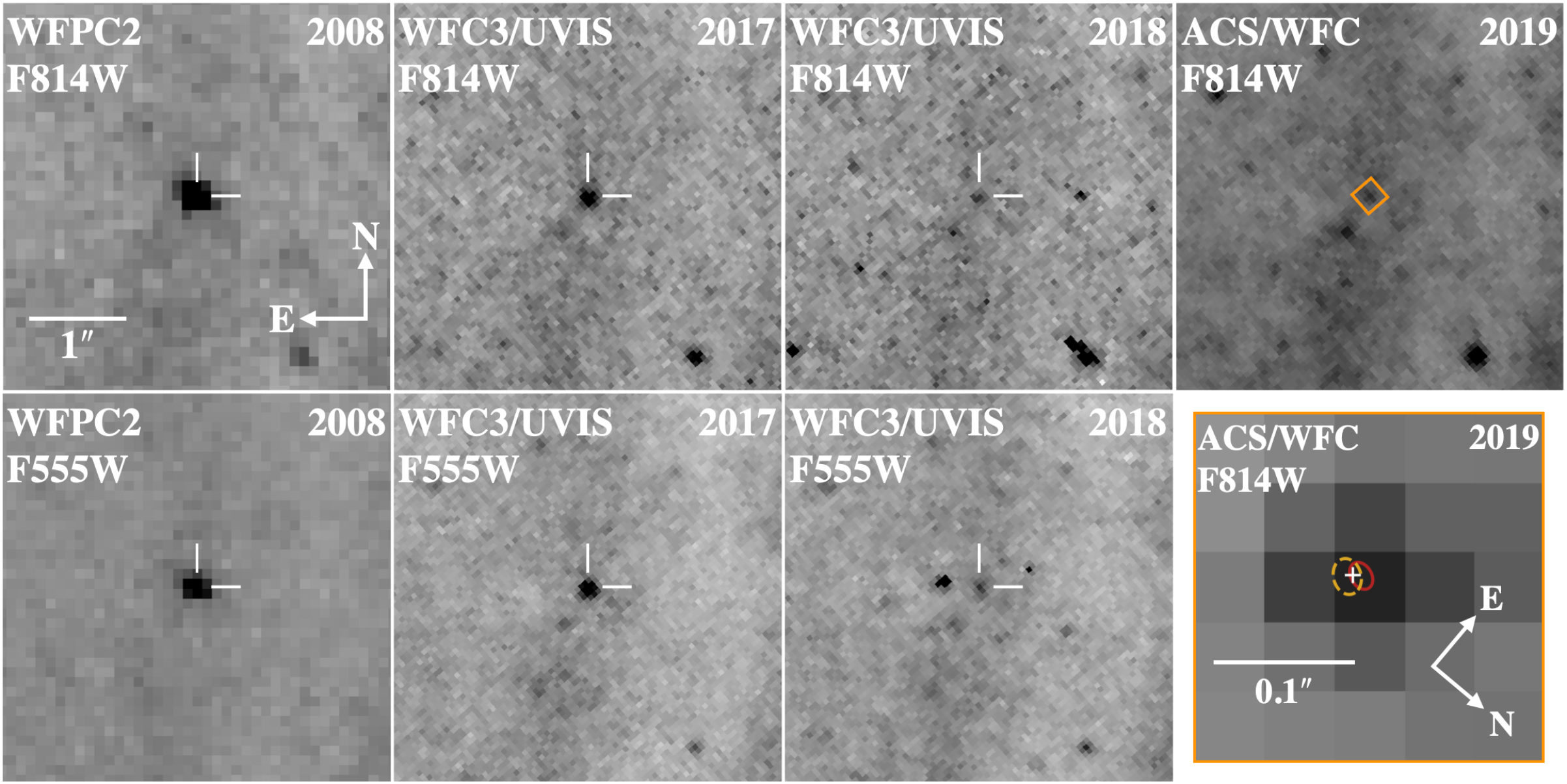}
\caption{\label{fig:hst_imaging}
HST imaging of the site of SN\,2015bh. In the top row, we show a sequence of available $F814W$ imaging, where the instrument and date of each image are indicated at the top of each panel and the location of SN\,2015bh indicated by the white crosshairs. The first three panels from the left in the bottom row show the same sequence in $F555W$. All of these panels are oriented N up and E left and are $4\arcsec$ on a side. The bottom, rightmost panel shows a zoom-in to a 5 pixel ($0\farcs25$) box around SN\,2015bh (indicated by the orange square in the upper, rightmost panel) in the 2019 $F814W$ image. The 1$\sigma$ confidence regions on the position of SN\,2015bh measured from the 2017 $F814W$ and $F555W$ images are shown as the red solid ellipse and yellow dashed ellipse, respectively  (see Appendix~\ref{appendix:astrometry}). The position of the source detected by \texttt{DOLPHOT} in the 2019 $F814W$ image, indicated by the white cross, is consistent with both measurements of the position of SN\,2015bh. 
}
\end{figure*}

In this paper, we confront these predictions with photometric data from HST taken between 2017 and 2019, nearly four years after the peak of the explosive transient. Following \citet{thone17}, we adopt the time of the observed peak on UT 2015 May 24.28 (MJD 57166.28) as the reference point for the phase of the transient ($t=0$~days) throughout. We also assume a total (Milky Way and host) extinction of $E(B-V) = 0.23$\,mag and employ the reddening law of \citet{fitzpatrick99} with $R_V = 3.1$ (see Appendix~\ref{appendix:host}). Our main result is that the source has faded well below the level of the progenitor in postexplosion, broadband imaging (see Figure~\ref{fig:hst_imaging}). SN\,2015bh joins SN\,2009ip \citep{smith22} and AT\,2016jbu \citep{brennan22c} in this respect, likely pointing to a terminal explosion, as has also been seen for at least some other Type~IIn SNe, such as SN\,1961V \citep{patton19} and SN\,2005gl \citep{gal-yam09}. 

%SN\,2015bh thus joins a small number of Type~IIn SNe whose massive progenitors appear to have faded in post-explosion imaging, including SN\,1961V (see above), SN\,2005gl \citep{gal-yam09} and SN\,2009ip itself \citep{smith22}. 
%\N{(N: and technically, also SN 1961V - now more than 7 mag fainter than progenitor....)}%Do I need a Section by section overview?
%In Section~\ref{sec:obs} we describe the observations, including newly present HST imaging and photometry (Section~\ref{sec:phot}) and a deep, late-time optical spectrum (Section~\ref{sec:spec_obs}). In Section~\ref{sec:results}, we describe our analysis to confirm the positional association of the faint residual emission with the transient (\ref{sec:astrometry}) and characterize the photometric evolution (Section~\ref{sec:lcs}) and the observed emission-line spectrum of the site (Section~\ref{sec:spec}). We discuss the possible interpretations for the source of the residual fading emission and the ultimate fate of the progentior in Section~\ref{sec:discussion}. Finally in Section~\ref{sec:summary}, we provide a brief summary of our results and main conclusions.

\section{Observations}\label{sec:obs}
Here, we present late-time HST imaging and photometry of SN\,2015bh, along with a deep optical spectrum at the transient location, in order to track and constrain associated emission as it fades below the progenitor level. All of the HST data presented in this paper were obtained from the Mikulski Archive for Space Telescopes (MAST) at the Space Telescope Science Institute. The specific observations analyzed can be accessed via \url{https://doi.org/10.17909/ksn7-y471}.

\subsection{Late-time HST Imaging}\label{sec:phot}
\begin{figure*}
\centering
\includegraphics[width=\textwidth]{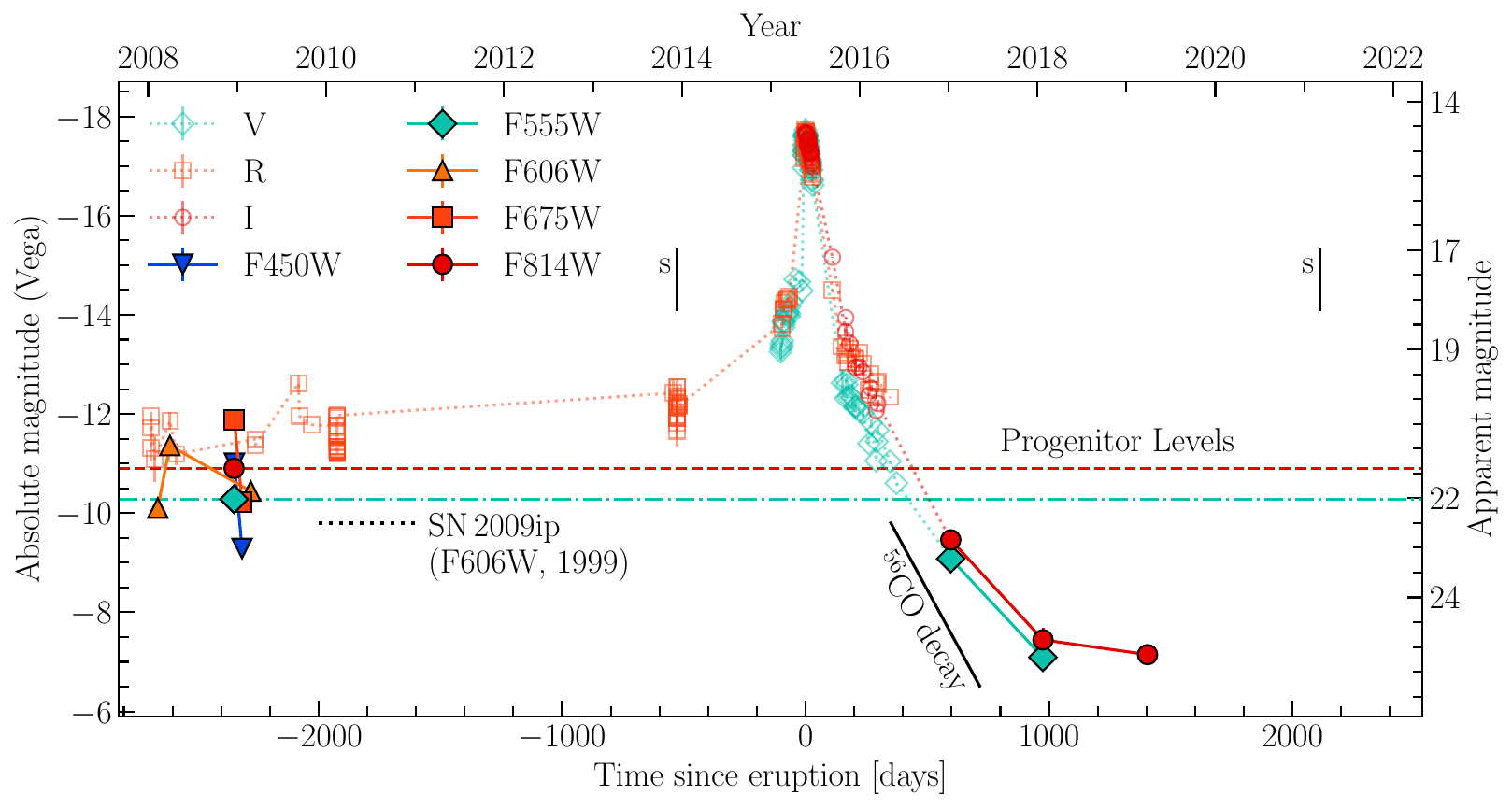}
\caption{\label{fig:hst_lcs}
The long-term pre- and postexplosion light curves of SN\,2015bh, %\edit1{
corrected for foreground extinction with $E(B-V) = 0.23$~mag. %}. 
HST photometry, including archival data from \citet{elias-rosa16} and newly presented late-time imaging, are shown as filled symbols. The late-time photometry has faded far below the observed levels of the progenitor at $F555W$ (dotted--dashed cyan line) and $F814W$ (dashed red line) in the 2008--2009 pre-explosion data. The 1999 progenitor level of SN\,2009ip at $F606W$ (dotted black line) is also shown for reference, %\edit1{
corrected for extinction with $A_R = 0.05$\,mag as in \citet{smith10}. %}. 
Ground-based $VRI$-band light curves of SN\,2015bh from \citet{goranskij16}, \citet{ofek16}, and \citet{elias-rosa16} are shown as open symbols. Spectroscopic observation epochs during the 2013 precursor eruptions (\citealp{ofek16}, see also Figure~\ref{fig:SEDs}) and our new late-time observations are indicated by vertical black bars and ``S'' symbols. 
%\textit{[Should I add lightcurves (say r-band) of other objects to this?]}.
}
\end{figure*}

The location of SN\,2015bh has been imaged by HST multiple times since the main SN-like event in 2015. The available late-time data include imaging with the Wide Field Camera 3 (WFC3) UVIS imager in the $F555W$ and $F814W$ filters taken on 2017 January 9.6, the $F438W$ and $F625W$ filters on 2017 February 17.9, and again in $F555W$ and $F814W$ on 2018 January 23.2 (PI: A.\ Filippenko, PIDs: 14668, 15166). Additional deep imaging with the Advanced Camera for Surveys (ACS) Wide Field Channel (WFC) in the $F814W$ filter was obtained on 2019 March 28.7 as part of an HST search for disappearing massive stars as failed SNe that form black holes \citep[PI D.\ Sand, PID 15645;][]{jencson22}. %\edit1{
We describe the image processing steps and use of the \texttt{DOLPHOT} package \citep{dolphin00,dolphin16} to extract photometry of SN\,2015bh in Appendix~\ref{appendix:phot}. We also confirm the positional association of the faint source detected in 2019 with SN\,2015bh with an astrometric analysis (Appendix~\ref{appendix:astrometry}). In the unlikely case of a chance coincidence with an unrelated source ($\lesssim$0.03\%), then the magnitudes reported here give upper limits on the SN or any remnant, as there are no other plausible counterparts in the vicinity. %}
%We examine the positional offsets of this source from the $F555W$ and $F814W$ 2017 catalog positions of SN\,2015bh in Figure~\ref{fig:dolphot_offsets}. In comparison to all other sources in the catalogs within a 500-pixel box around SN\,2015bh and with comparable $S/N \geq 15$, the position of the 2019 source is consistent with those of the SN\,2015bh detections at $\lesssim$1$\sigma$ in both filters. 
Our photometry (Vega magnitudes) is presented in Table~\ref{table:HST_phot} and the long-term, multiband light curves are shown in Figure~\ref{fig:hst_lcs}. 

%To confirm the positional association of the faint source detected in 2019, we perform a %separate, more
%detailed astrometric analysis and assess the probability that the 2019 source is a chance coincidence with an unrelated object in Appendix~\ref{appendix:astrometry}. If the source is unrelated, then the magnitudes reported here give upper limits on the SN or any remnant, as there are no other plausible counterparts in the vicinity. 
%\N{(N: this is sorta obvious, but you could also note that regardless of positional accuracy, there is no other bright source nearby that could be the SN or progenitor or companion...so if this isn't the source, than these magnitudes give upper limits to its luminosity.)}

\begin{deluxetable*}{ccccc}
\tablecaption{HST \texttt{DOLPHOT} Photometry \label{table:HST_phot}}
\tablehead{\colhead{UT Date} & \colhead{MJD} & \colhead{Inst.} & \colhead{Band} & \colhead{App.\ Magnitude\tablenotemark{a}} %\\ 
%\colhead{} & \colhead{} & \colhead{} & \colhead{} & \colhead{(mag)}
}
\startdata
2017 Jan 09.59 & 57762.59 & WFC3/UVIS & $F814W$ & 23.24 (0.11) \\
2017 Jan 09.60 & 57762.60 & WFC3/UVIS & $F555W$ & 23.95 (0.06) \\
2017 Feb 17.88 & 57801.88 & WFC3/UVIS & $F438W$ & 24.47 (0.15) \\
2017 Feb 17.88 & 57801.88 & WFC3/UVIS & $F625W$ & 23.37 (0.08) \\
2018 Jan 23.21 & 58141.21 & WFC3/UVIS & $F814W$ & 25.25 (0.24) \\
2018 Jan 23.22 & 58141.22 & WFC3/UVIS & $F555W$ & 25.94 (0.14) \\
2019 Mar 28.69 & 58570.69 & ACS/WFC & $F814W$ & 25.55 (0.20) \\
\enddata
\tablenotetext{a}{%\edit1{
Observed Vega magnitudes. No extinction corrections applied. %} 
1$\sigma$ uncertainties are given in parentheses.}
\end{deluxetable*}

%\begin{figure}
%\centering
%\includegraphics[width=\columnwidth]{NGC2770-SN2015bh_WFC3_UVIS_F555W+F814W_ACS_WFC_F814W_Dolphot_offsets_preproc.pdf}
%\caption{\label{fig:dolphot_offsets}
%Positional offsets between the 2017 WFC3/UVIS imaging ($F555W$ left, $F814W$ right) and the 2019 $F814W$ imaging of matched stars in our \texttt{DOLPHOT} catalogs ($S/N \geq 15$) within a 500-pixel box centered on the location of SN\,2015bh (gray crosses). %The subset of these stars that appear relatively isolated upon visual inspection of the images are shown in black. 
%The 1- and 2$\sigma$ ellipses for the stars are shown as the gray dashed and dotted curves, repsectively, assuming the offsets follow a 2D Gaussian distribution. SN\,2015bh itself in 2017 and the nearest source detected in 2019 (red stars) are positionally coincident at $\lesssim$1$\sigma$ in both filters.}
%\end{figure}

\subsection{Spectroscopy}\label{sec:spec_obs}
\begin{figure*}
\centering
\includegraphics[width=\textwidth]{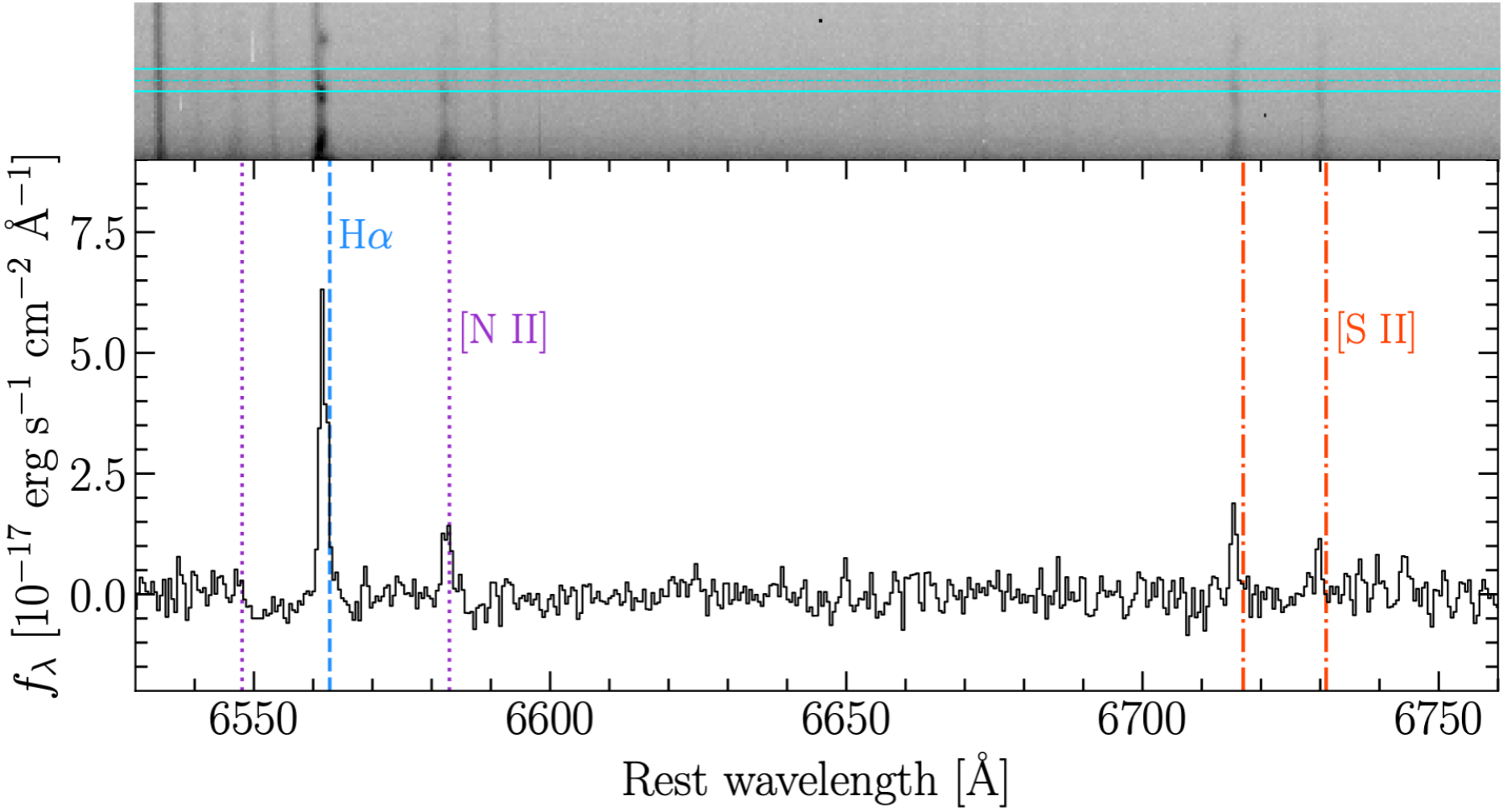}
\caption{\label{fig:spec}
Late-time MMT Blue Channel spectrum of the site of SN2015bh in the rest frame of NGC\,2770 %\edit1{
($z=0.00649$, from NED). %}. 
The extraction aperture is shown overlaid on the reduced 2D spectrum at the top of the figure. %\edit1{
Rest-frame wavelengths of %} 
emission features typical of \ion{H}{2} regions are indicated by the vertical lines and labeled by atomic species. The spectrum has been dereddened only for Milky Way extinction to NGC\,2770. The observed lines are at a velocity of $-40$\,km~s$^{-1}$ compared to the systemic velocity of NGC\,2770, consistent with the galaxy's rotation curve \citep{marquez02}.} %\N{N: offset = galaxy rotation.  is there a reference for the galaxy's rotation curve?}}
\end{figure*}

We obtained a moderate-resolution optical spectrum on 2021 March 7.3, more than 5~yr after the explosion, with the Blue Channel Spectrograph at MMT Observatory on Mount Hopkins in Arizona. We used a 1200\,l~mm$^{-1}$ grating with a central wavelength of 6362~\AA\ and a $1\farcs0$ slit. This provides a wavelength coverage of $5700$--$7000$~\AA\ and a resolving power of $\mathcal{R}=4500$. As SN\,2015bh was faint at this phase (see light curves in Figure~\ref{fig:hst_lcs}), we first obtained a short 120\,s exposure on a nearby reference star. We then performed a blind offset to acquire the position of SN\,2015bh, where we obtained a total of $5 \times 1200$\,s exposures. Standard reduction procedures were carried out using IRAF \citep{tody86}. We used the reference star observation to determine the position of SN\,2015bh on the slit and the shape of the spectral trace for extraction. We performed flux calibration of the 1D spectrum using observations of a spectrophotometric standard taken at a similar airmass on the same night. The reduced spectrum in shown in Figure~\ref{fig:spec}. 
%The spectrum, shown in Figure~\ref{fig:spec}, appears consistent with that of an \ion{H}{2} region. Though we do not detect any clear features of the SN (e.g., broad lines), we present an analysis of the spectrum in Section~\ref{sec:spec}.

\section{Analysis and Results} \label{sec:results}

\subsection{Late-time Photometric Evolution}\label{sec:lcs}
The long-term evolution in the light curves of SN\,2015bh is shown in Figure~\ref{fig:hst_lcs}, from the earliest HST detections of the progenitor star in 2008--2009 \citep{elias-rosa16}, through several precursor eruptions, the multiphase main outburst in 2015 \citep{goranskij16,ofek16,elias-rosa16}, and the very late-time fading of the SN in our newly presented HST photometry extending nearly four years after the explosion. % (see also Table~\ref{table:HST_phot}). 
A comprehensive photometric history of the source up to 2016, including precursor variability observed since 1994 is also presented in \citet{thone17}. The primary result of the present work is that between 2017 and 2019 the SN has faded significantly below the level of the 2008 progenitor detections in broadband filters, notably by $3.8 \pm 0.2$~mag at $F814W$ as of March 2019.

Between 2017--2018 ($t=596$--975\,days), the source fades at a rate of $0.0053 \pm 0.0004~(\pm0.0007)$~mag~day$^{-1}$ at $F555W$ ($F814W$). There is essentially no observed color evolution between $F555W$ and $F814W$ during this time, with $F555W - F814W = 0.4 \pm 0.1~(\pm0.3)$ at $t=596~(975)$\,days. This is somewhat bluer than the latest ground-based photometry reported by \citet{thone17} in 2016 at a phase of $t=239$\,days in comparable filters of $V-I = 1.2 \pm 0.4$~mag and bluer than the 2008 progenitor source at $F555W - F814W = 0.62 \pm 0.04$~mag \citep{elias-rosa16}. Between 2018 and the latest observation in 2019 ($t=975$--1404\,days), the fade rate at $F814W$ slows to $0.0007 \pm 0.0007$~mag~day$^{-1}$, nominally consistent with a flat evolution. Correcting for foreground extinction, the source reaches its faintest ever observed magnitude of $F814W = 25.1 \pm 0.2$~mag ($M_{F814W} = -7.1$~mag) in the final $F814W$ observation. 

\subsubsection{Spectral Energy Distribution Evolution}\label{sec:SED}
\begin{figure}[htb!]
\centering
\includegraphics[width=\columnwidth]{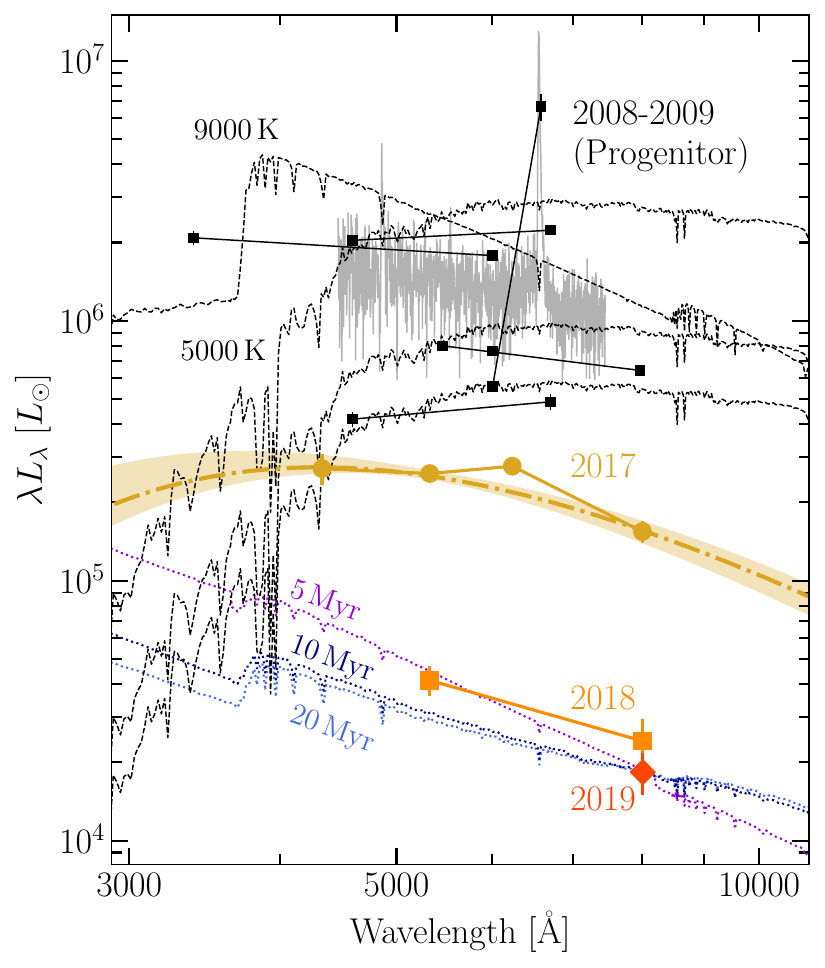}
\caption{\label{fig:SEDs}
SEDs from HST photometry of SN\,2015bh in 2017 (yellow circles), 2018 (orange, thin diamonds), and 2019 (red, thick diamond). The best-fit blackbody spectrum ($T_{\mathrm{BB}} = 8530$~K, $R_{\mathrm{BB}} = 280$~$R_{\odot}$) to the 2017 data (excluding $F625W$) is shown as the yellow dashed curve and the yellow shaded regions represents the 16th--84th percentile uncertainties from the MCMC. %A blackbody spectrum of the same temperature, but scaled to the 2018 photometry ($R_{\mathrm{BB}} = 100$~$R_{\odot}$) is shown as the orange, dashed curve. %%JJ: I cut this cause it clutters the plot and isn't super useful. 
The 2008--2009 HST photometry of the progenitor from \citet{elias-rosa16} is shown as black squares, where the lines connect contemporaneous points. The black dotted curves show ATLAS synthetic stellar spectra \citep{castelli03} with the properties inferred for the progenitor listed in Table~4 of \citet{elias-rosa16}. The observed 2013 spectrum of the progenitor from \citet{ofek16} is shown in gray. Finally, the dotted purple, indigo, and blue curves represent model Starburst99 clusters at ages of 5, 10 and 20~Myr, respectively, as described in the main text. 
}
\end{figure}

We constructed multiepoch spectral energy distributions (SEDs) of SN\,2015bh from the available pre- and postexplosion HST photometry, as shown in Figure~\ref{fig:SEDs}. The photometric magnitudes were converted to band luminosities using the filter zero-points and effective wavelengths available in the \texttt{pysynphot} package \citep{pysynphot}. %\J{JS: Do you mean $\lambda L_{\lambda}$, or do you mean the luminosity integrated over the band? These aren't quite the same thing.}. 
We reproduce the characterization by \citet{elias-rosa16} of the 2008--2009 progenitor photometry with ATLAS synthetic stellar spectra \citep{castelli03} with effective temperatures $T_{\mathrm{eff}} = 5000$--$9000$~K and bolometric luminosities $\log L = (3$--$15) \times 10^{39}$~erg~s$^{-1}$ ($\log[L/L_{\odot}] = 5.9$--$6.6$). %\edit1{
This characterization is mostly illustrative, as the observed 2013 progenitor showed strong emission lines indicative of intense mass loss \citep{ofek16,boian18}.%}

In 2017 ($t = 596.3$ and 635.6\,days), the source has already faded well below any of the available progenitor photometry in broadband optical filters. We fit a blackbody spectrum to the data, excluding the $F625W$ point that we expect is contaminated by H$\alpha$ emission. We performed a Markov chain Monte Carlo (MCMC) simulation with the  \texttt{lightcurve\_fitting} package \citep{hosseinzadeh_lc_zenodo}\footnote{\url{https://griffin-h.github.io/lightcurve_fitting/index.html}}, from which we adopt the 16th and 84th percentiles of the posterior distributions as estimates of the uncertainties in the fitting parameters. We obtained $T_{\mathrm{BB}} = 8530_{-590}^{+1310}$~K and $R_{\mathrm{BB}} = 280_{-50}^{+40}$~$R_{\odot}$, corresponding to a luminosity of $L_{\mathrm{BB}} = 1.45_{-0.06}^{+0.23} \times 10^{39}$\,erg~s$^{-1}$ ($\log[L/L_{\odot}] = 5.57_{-0.02}^{+0.07}$). %\edit1{
In 2018, the blackbody temperature of the source is poorly constrained with $F555W$ and $F814W$ measurements alone (the MCMC yields $T_{\mathrm{BB}} \approx 8000$--44,000~K). Requiring that $L_{BB}$ in 2018 be lower than the 2019 upper bound, however, constrains $T_{\mathrm{BB}} \lesssim$ 20,000~K in 2018. Still, given the lack of observed color evolution, it seems unlikely that the SED has changed significantly. Simply scaling the 2017 blackbody with $T_{\mathrm{BB}} = 8530$~K to the 2018 photometry, we find $L_{\mathrm{BB}} \approx 2.3 \times 10^{38}$\,erg~s$^{-1}$ ($\log[L/L_{\odot}] = 4.8$). This is a factor of $\approx$13 fainter in luminosity than the faintest level inferred from the 2008--2009 progenitor photometry. These estimates should be viewed with additional caution, as the spectrum of the fading source is likely dominated by emission features related to ongoing CSM interaction (see Section~\ref{sec:csm}), in which case a blackbody approximation may not be appropriate.%}
%\N{(N: yeah, the lack of color is the main thing.  all that stuff about fitting the temperature is not very meaningful and maybe a little misleading, since you know it isn't a blackbody anyway - you can fit a blackbody to 2-filter broadband images of an Hii region, but the temperature you get is nonsense...)} \N{(N: pretty sure that R$_{BB}$ and $L_{BB}$ are meaningless, but ok...)} 

%It remains possible, however, that a hot, more luminous source is still present below the level of the observed 2018--2019 $F555W$ and $F814W$ photometry. To illustrate this in Figure~\ref{fig:SEDs}, we show ATLAS stellar spectra of $T_{\mathrm{eff}} = 30,000$, 40,000, and 50,000~K with $\log[L/L_{\odot}] = 5.9$, the same bolometric luminosity as inferred from the HST photometry of the progenitor at its faintest-observed, pre-explosion level. The 2018 $F555W$ measurement constrains the temperature of any such source to $>$30000~K. We discuss this scenario further in the context of a possible surviving remnant star in Section~\ref{sec:star}. 

\begin{figure*}[htb!]
\centering
\includegraphics[width=\textwidth]{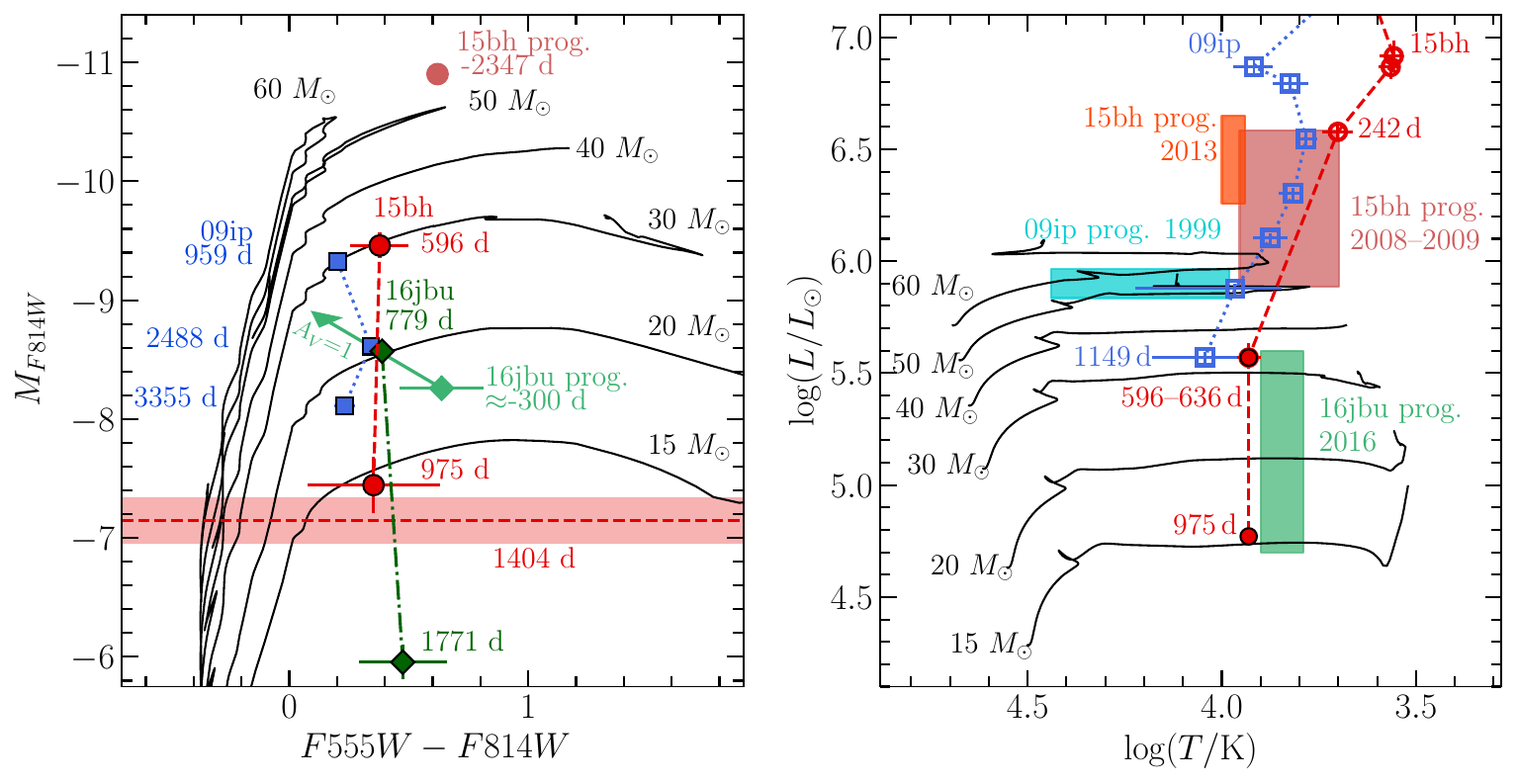}
\caption{\label{fig:cmd_hrd}
Left: CMD from the $F555W$ and $F814W$ imaging of SN\,2015bh, including a detection of the progenitor in 2008 (light-red circle; \citealp{elias-rosa16}) and new late-time points from 2017 to 2019 (red circles). Late-time points for SN\,2009ip \citep[blue squares;][]{smith16b,smith22} and AT\,2016jbu \citep[dark-green diamond;][]{brennan22b} are also shown. For the AT\,2016jbu progenitor (light-green diamond), the arrow represents the $A_V = 1$~mag reddening vector \citep{kilpatrick18a}. All phases are relative to the transient peak.  Right: HRD depicting the late-time evolution of SN\,2015bh (red circles) and SN\,2009ip (blue squares). Unfilled points are based on ground-based data and reproduced from \citet[][see their Figure~15]{thone17}, while new points for SN\,2015bh are shown as filled symbols. %\edit1{
The evolution shown assumes a constant temperature between the last two epochs, but an evolution to higher temperatures $\lesssim$20,000~K with a smaller drop in luminosity at $t=975$\,days is also consistent with observations. %} 
The inferred properties of the progenitor stars are indicated by the shaded regions, including from the 1999 HST photometry of SN\,2009ip \citep[cyan;][]{smith10,foley11}, the 2016 HST photometry of AT\,2016jbu \citep[light green;][]{kilpatrick18a,brennan22b}, the 2008--2009 HST photometry of SN\,2015bh \citep[light red;][]{elias-rosa16}, and from modeling of the SN\,2015bh 2013 progenitor spectrum \citep[light red;][]{boian18}. %The properties of the source underlying the observed cool, optically thick wind in the models of \citet{boian18} are indicated by the red-orange, dotted-outline region. %%decided to cut this. Don't really talk about it.  
Nonrotating, solar-metallicity stellar evolution tracks from MIST \citep{choi16} are shown for comparison as black curves for a range of initial masses in both panels. 
}
\end{figure*}

\subsection{Late-time spectrum}\label{sec:spec}
As shown in Figure~\ref{fig:spec}, the late-time spectrum, taken at $t=2114.0$\,days, shows only narrow emission lines characteristic of \ion{H}{2} regions, namely H$\alpha$, [\ion{N}{2}] ($\lambda\lambda$ 6548, 6583), and [\ion{S}{2}] ($\lambda\lambda$ 6717, 6731). The observed width of the lines, $\mathrm{FWHM} = 80$\,km~s$^{-1}$, is consistent with the instrumental resolution, and we see no evidence of shock-broadened features associated with the SN. %The lines are at a  blueshifted velocity of $-40$\,km~s$^{-1}$ compared to the systemic velocity of NGC\,2770, consistent with the galaxy's rotation curve \citep{marquez02}. 
We constrain the integrated H$\alpha$ flux from the SN to $\lesssim\! 7\times10^{-16}$~erg~s$^{-1}$~cm$^{-2}$ (3$\sigma$), using the broadened ($\approx$2000~km~s$^{-1}$) double-peaked profile from the latest available spectrum at $t=242$\,days from \citet{thone17} as a template. This corresponds a limit on the broad H$\alpha$ luminosity of $\lesssim\! 7\times10^{37}$~erg~s$^{-1}$ (Galactic extinction only) or $\lesssim\! 10^{38}$~erg~s$^{-1}$ (host extinction included). This limit is conservative if the line profile narrows as the SN evolves.

In the 2D spectral image shown in Figure~\ref{fig:spec}, the narrow emission within our extraction aperture appears to be dominated by an extended H$\alpha$ clump, whose brightest point is $\approx$2$\arcsec$ below the expected position of SN\,2015bh on the slit ($\approx$280~pc projected separation). This may indicate that the emission arises primarily from a nearby, but likely not directly associated, \ion{H}{2} region. We perform additional analyses of this emission to characterize the nearby environment of SN\,2015bh in Appendix~\ref{appendix:spec}.

\section{Discussion}\label{sec:discussion}
As described above, SN\,2015bh has now faded far below the level of its well-characterized progenitor star in broadband optical filters. Here, we discuss the possibilities for the source of the fading emission and implications for the fate of the progenitor. 

\subsection{Ongoing Circumstellar Interaction}\label{sec:csm}
The previously published, late-phase spectra of SN\,2015bh ($t\approx130$--290~days) were dominated by strong H$\alpha$ emission with an irregular, double-peaked profile and velocities $\approx$2000~km~s$^{-1}$, along with emission features of [\ion{Ca}{2}], the \ion{Ca}{2} IR triplet, narrow ($\approx$500~km~s$^{-1}$) emission features of \ion{He}{1} and \ion{Fe}{1}, and the emergence of a pseudo-continuum blueward of 5450~\AA\ attributable to a forest of Fe emission lines. Altogether, these features indicate the continued influence of interaction with asymmetric CSM  \citep{elias-rosa16,thone17}. Strong H$\alpha$ emission and signs of CSM interaction at late phases appear to be ubiquitous in objects similar to SN\,2015bh, including SN\,2009ip through at least 726--1196\,days after peak \citep{fraser15,smith16b}, AT\,2016jbu up to 419\,days after peak \citep{brennan22a}, and SN\,2011fh up to 1359 days after peak \citep{thallis22}. As noted too by \citet{kilpatrick18a}, the double-peaked asymmetric H$\alpha$ profiles of AT\,2016jbu and SN\,2015bh at late times are particularly similar.

The light from SN\,2015bh in 2017 must be dominated by the explosion and related interactions, as it continued to fade significantly by the time of the 2018 observations. As discussed in Section~\ref{sec:lcs} and shown in Figure~\ref{fig:hst_lcs}, the source continued to fade through 2018 in $F555W$ and $F814W$ at a similar rate to the earlier ground-based photometry ($\approx$0.005~mag~day$^{-1}$). This is notably slower than the predicted rate for the decay of  $^{56}$Co, which, as pointed out by \citet{elias-rosa16}, likely points to the continued contribution of CSM interaction. %\edit1{
As shown in the CMD in Figure~\ref{fig:cmd_hrd}, the fading of SN\,2015bh at a constant color $F555W - F814W\approx 0.4$~mag is remarkably similar that of other objects in the SN\,2009ip-like class, namely, SN\,2009ip itself \citep{smith16b,smith22} and AT\,2016jbu \citep{brennan22c,brennan22b}. We note, too, that the late-time detections of all three sources are bluer than the available progenitor detections of SN\,2015bh and AT\,2016jbu at $F555W - F814W\approx 0.6$~mag, though as noted by \citet{kilpatrick18a}, the progenitor of AT\,2016jbu may have suffered circumstellar extinction at the level of $A_V \approx 1$~mag. %} 
This color may reflect the numerous emission lines produced by ongoing CSM interaction in the spectra of these events, specifically the Fe pseudo-continuum blueward of 5450~\AA\ that contributes to the $F555W$ flux, but also the \ion{Ca}{2} IR triplet emission in the $F814W$ band. 
%\N{(N: but probably also CaII IR triplet in F814W.)} g

Between 2018 and 2019, the $F814W$ light curve flattens out considerably. If the light curve is still primarily powered by CSM interaction, this may indicate a change in the density profile of the CSM. While we do not see any signs of interaction ($L_{\mathrm{H}\alpha} \lesssim 10^{38}$~erg~s$^{-1}$ for broadened emission; Section~\ref{sec:spec}) in our 2021 optical spectrum at $t=2114$ days (710~days after the last $F814W$ image), %\J{JS: how constraining is this? you do have an estimate of the H-a lum or limit}, 
these observations do not constrain the presence of interaction at earlier phases. We discussion alternative scenarios to explain the slowing decline rate below, including contributions from a scattered-light echo (Section~\ref{sec:echo}) or the settling of the light curve onto a persistent, quiescent source, i.e., a possible binary companion or host stellar cluster (Section~\ref{sec:cluster}), or (less likely) the surviving remnant of the progenitor (Section~\ref{sec:star}).

\subsection{Scattered-light Echo}\label{sec:echo}
%We now consider whether the flattening of the $F814W$ light curve between 2017 and 2019 may be attributed to an unresolved light echo. 
In the light-echo scenario, additional light from the transient is scattered by circumstellar or more distant dust into the line of sight of the observer. The echo has a time delay because of the longer light path compared to direct, unscattered light from the transient. For SNe with CSM from recent pre-explosion mass loss, a scattered-light echo from circumstellar dust will lead to a flatting of the optical light curve and is also expected to accompany a thermal echo, as the heated dust will reradiate light absorbed from the transient into the IR \citep{chevalier86}. This argument has been used, for example, to explain the late-time optical emission of the CSM-interacting SN\,2006gy \citep{smith08,miller10,fox15b}. Scattered-light echoes may also arise from a foreground distribution of dust, as claimed, e.g., in the cases of the Type IIb SN\,2011dh \citep{maund19} and the Type Ia SN\,1998bu \citep{cappellaro01}. The SED of a scattered-light echo will resemble the luminosity-weighted average of the time-varying SED of the transient, which will be dominated by the spectrum of the transient at peak. An echo may also appear bluer owing to the enhanced scattering efficiency of dust grains at shorter wavelengths.
%\N{(N: although a scattered echo is also likely to be even a little bluer than the SN peak, since the scattering efficiency induces a blue tilt unless the grains are huge...)}

At peak, the optical colors of SN\,2015bh are relatively blue at $V-I = -0.06 \pm 0.04$~mag \citep{elias-rosa16}. In 2017 and 2018, the comparable $F555W - F814W$ color stays largely constant at $\approx$0.4~mag, though the uncertainty in 2018 is larger ($\pm$0.3~mag). This argues against an echo %, which should reflect the bluer color of the transient peak, 
as the dominant source of the emission in 2018, %\edit1{
though it cannot be conclusively ruled out as the SED of the echo will depend on the grain properties of the dust and its geometry \citep[see, e.g.,][]{maund19}. %}. 
We lack color information to derive further constraints on the echo scenario in 2019, but note that our 2021 spectrum does not display any features (e.g., broad H$\alpha$) reminiscent of the transient at peak. Lastly, the 2019 source is well fit by an ACS/WFC PSF, implying a projected size $\lesssim$0.5 WFC pixels or $\lesssim\! 0\farcs02$--$0\farcs03$. An echo will become more obvious as a spatially extended source with time, a scenario that could be tested with future HST imaging.  

%In 2019, we have only an $F814W$ image, but were the source flux at $F555W$ to remain constant, the bluest the source could be is $F555W - F814W \approx 0.0 \pm 0.2$~mag. This is closer to the transient color at peak, and thus it seems plausible that the flattening of the $F814W$ light curve between 2018 and 2019 could point to a significant contribution from an optical scattered light echo. This scenario may be testable with continued multiband monitoring of the source. If true, this would imply that any contributions from an underlying companion, cluster, or the surviving progenitor would be even fainter. 
%\N{(N: but also, if a scattered echo dominated, the spectrum would look like the peak of the SN, with broader Halpha.)}
%\J{JS: You should also mention the good PSF fit for a single star to the data (i.e. that a PSF fits well at late times). Without extra sims I think it's hard to *exactly* say what the limit on the size is since it's S/N dependent (people regularly do 0.02" with ACS, but 0.008-0.01" is harder), plus of course the light echo predictions are geometry dependent. But I think a broad statement that you see no evidence for a spatial extent of the source in 2019, and that a light echo would be expected to become more obvious with time and hence can be tested with future data, is basically model independent and safe.}  

%peak V-I = -0.06 +- 0.05 #ER16
%color 2008: 0.62 +- 0.04
%      2017: 0.38 +- 0.12
%      2018: 0.35 +- 0.28
%      2019: >0.054 +- 0.24

\subsection{Binary companion or Stellar Cluster}\label{sec:cluster}
In the absence of a light echo, the flattening of the $F814W$ light curve may suggest that SN\,2015bh is fading below the level of a persistent source. If we assume a constant rate of decline (in magnitudes) of the CSM-interaction-powered source (see section~\ref{sec:csm}), the late-time $F814W$ light curve would imply a constant underlying source at $M_{F814W} = -7.0$~mag. This is consistent with the 2019 brightness at $M_{F814W} = -7.1\pm0.2$, suggesting that the photometry at this phase is dominated by the flux of the persistent source. 
%%%I might move this analysis to the light curve section and reference it here. I think it's clear what I did, but maybe I should show a figure showing the constant+fading components? 

One possibility is that the flux of the persistent source arises primarily from the binary companion of the progenitor. This may be expected given the proposed binary origins of SN\,2015bh and has been claimed previously for two stripped-envelope SN~IIb (SN\,1993J, \citealp{maund04,fox14}; SN\,2011dh, \citealp{folatelli14,maund19}). The CMD position in 2018 ($t=975$\,days) is similar to that of a $\approx$15~$M_{\odot}$ star evolving across the Hertzsprung gap (Figure~\ref{fig:cmd_hrd}). %Finding a companion star at an identical color to the 2017 interaction-powered source is unlikely, 

An evolving 15~$M_{\odot}$ star will spend $\lesssim$10$^{5}$~yr in the Hertzsprung gap and $\lesssim$10$^{4}$~yr the observed CMD position, compared to its main-sequence lifetime of $\approx$10$^{7}$~yr (\citealp{choi16}; see tracks in Figure~\ref{fig:cmd_hrd}).
%This would be inconsistent with the believed \N{(N: not sure i believe this.  maybe say "claimed" or "proposed")} high mass ($\gtrsim35$~$M_{\odot}$) of the progenitor, as a lower mass companion would have a much longer main-sequence lifetime. \J{JS: I agree with Nathan, seems easy to get around these vague constraints with an outburst implies lower mass progenitor + various binary evolution scenarios. A timing argument (that an H-gap star spends virtually zero time at this color) is probably stronger.}
%\edit1{
For most SNe then, %} 
a companion is more likely to still be on the main sequence as an O- or B-type star. Any such star $\lesssim$30~$M_{\odot}$ would have $M_{F814W} \gtrsim -6$~mag, too faint to contribute much flux to the persistent $F814W$ source inferred above. A more massive O-type companion at $M_{F814W} = -7.1$~mag would be too blue with $M_{F555W} \approx -7.4$~mag, inconsistent with the 2018 measurement at $M_{F555W} = -7.1$~mag. %\edit1{
Still, a secondary star may plausibly appear in the Hertzsprung gap for binary mass ratios close to one (see, e.g., SN\,1993J, \citealp{maund04}; SN\,2006jc, \citealp{sun20}; SN\,2019yvr \citealp{sun22}). Alternatively, a main-sequence companion may appear cooler and more luminous if it is temporarily inflated by interaction with the SN ejection \citep[e.g.,][]{hirai18}.%}

%As discussed below, it is possible that the residual flux is instead from a low-mass cluster or star-forming region.%Moreover, massive stars and binaries are rarely born in isolation, and it is much more likely that the persistent flux arises from an associated cluster or star-forming region. \J{JS: I feel like this statement is akin to saying that most of SN progenitors have an associated $M_V=-7.2$ stellar population which I don't think is true, or at least would need additional justification/citations. I'd just say that it's instead *possible* that the residual flux is from a low-mass cluster or star forming region.}

It is also possible that the residual flux is from low-mass cluster or star-forming region. In Figure~\ref{fig:SEDs}, we show simulated Starburst99 \citep{leitherer99} star cluster SEDs\footnote{Starburst99 simulations can be generated at \url{https://www.stsci.edu/science/starburst99/docs/default.html}} at ages of 5, 10, and 20\,Myr, scaled to this level at $F814W$. These correspond to main-sequence turnoff masses of 54, 21, and 12~$M_{\odot}$ and luminosities $M_V = -7.1$, $-6.7$, and $-6.6$~mag. %, and luminosity-scaled cluster masses of 4, 2.3, and 1.8~$M_{\odot}$, respectively. The cluster masses should not be interpreted literally, as the base assumption that the initial mass function (IMF) is fully populated in the Starburst99 models is clearly violated; however 
This suggests that the light from the persistent source may be dominated by a relatively small number of the brightest stars.\footnote{Proper treatments of low-mass clusters, where stochastic effects in sampling the IMF are important, can be performed with, e.g., \texttt{SLUG} (Stochastically Lighting Up Galaxies; \citealp{dasilva12}).} We have implicitly assumed that the potential cluster suffers the same host extinction as SN\,2015bh itself ($E[B-V]_{\mathrm{host}} = 0.21$~mag); if the average extinction to the cluster were lower, its inferred luminosity would be lower as well. The persistent source is well fit with an ACS/WFC PSF (see Section~\ref{sec:phot}), implying a projected size $\lesssim$0.5 WFC pixels or $\lesssim$3\,pc at the assumed distance to NGC\,2770.
%\J{JS: people regularly measure faint star clusters with radii of ~ 2 pc at the distance of Virgo, so ~ 3 pc at N2770. It depends slightly on brightness, but the answer isn't too far from ~ 3 pc.}
%3.4 pc is 0.5 ACS pixels, this seems reasonable. 
%as Jay said, maybe we can do better then this?
%

Based on the latest photometry of SN\,2009ip up to 3355 days postexplosion, \citet{smith22} suggest the presence of a cluster with $M_V \approx -7.5$~mag, possibly similar in scale to Trumpler 14, a $\sim$1--2~Myr-old, $4300~M_{\odot}$ cluster in the Carina Nebula \citep{vazquez96}. They infer an older age of $\gtrsim$4--5~Myr, however, based on the lack of a bright, resolved H$\alpha$ nebula that would be associated with a younger cluster. Our constraints on the host cluster of SN\,2015bh ($M_V \gtrsim -7.1$) would then imply a smaller, less massive association. For example, the Trapezium cluster ($M_V \approx -6$~mag) within the $\approx$1800~$M_{\odot}$, $\approx$2~Myr-old Orion Nebula hosts a single O-type star of mass $M\approx 30~M_{\odot}$ \citep{hillenbrand98}; the host cluster of SN\,2015bh may plausibly lie somewhere in between these two examples. The luminosity of the nearby ($\lesssim$300\,pc) H$\alpha$ clump of $L_{\mathrm{H}\alpha} = 2$--$3\times10^{37}$~erg~s$^{-1}$ (Appendix~\ref{appendix:spec}) is in the range of classical (e.g., Orion; 10$^{37}$~erg~s$^{-1}$) up to giant star-forming regions (e.g., SMC N66; $6\times10^{38}$~erg~s$^{-1}$; \citealp{crowther13}). In contrast, SN\,2009ip is largely isolated from other signs of star-formation \citep{smith16b,smith22}. Narrowband H$\alpha$ imaging would provide valuable insight to the spatial morphology of this region and any possible emission associated with SN\,2015bh. Now that SN\,2015bh appears to have faded below the level of a persistent source, multiband photometry from the UV to the IR can %\edit1{
test for the presence of an evolved or inflated companion %} 
and probe the age and stellar content of the host population. This will provide important constraints on evolutionary pathways that could have produced the progenitor system.

\subsection{Did the Progenitor Star Survive?}\label{sec:star}
The late-time data presented here show that SN\,2015bh is now $\approx$4~mag fainter than its eruptive progenitor in optical filters and rule out a scenario in which the source has simply returned to its pre-explosion state. An essential question is whether the star survived the explosion, but is now in a dramatically altered state (e.g., hot and optically faint), or if it is truly gone. The most important new clues are (1) the substantial fading of the source below the observed level of the progenitor in broadband optical filters (by nearly 4~mag at $F814W$), and (2) the source is observed to fade in the $F555W$ and $F814W$ bands at identical rates. %\edit1{
Importantly, this strongly indicates that the optical fading is not the result of postshock dust formation, which is seen in some Type IIn supernova \citep[e.g.,][]{jencson16} and would cause the source to appear redder with time. %} 
This is extremely similar to the late-time evolution of SN\,2009ip recently presented by \citet{smith22}, and we come to similar conclusions about the fate of the progenitor of SN\,2015bh.

The substantial fading of SN\,2015bh below its progenitor level is easily explained if the progenitor star has simply vanished. The pre-explosion observations of the progenitor, however, indicate a luminous, highly variable star in outburst (see Section~\ref{sec:SED} and discussions of the progenitor in, e.g., \citealp{elias-rosa16,thone17}), and it is important to consider that the quiescent star may be fainter. \citet{boian18}, for example, infer a progenitor mass $\gtrsim$35~$M_{\odot}$ from their detailed modeling of the 2013 progenitor spectrum as an optically thick wind---if the observed luminosity of the star is at the Eddington limit in outburst---but they do not rule out the possibility of an lower-mass, intrinsically fainter progenitor with a dynamic super-Eddington wind \citep[see, e.g.,][]{shaviv01,owocki04,vanmarle09}. %As shown in Figure~\ref{fig:cmd_hrd}, the color and inferred luminosity of SN\,2015bh at $>$975\,days is well below that of a $35~M_{\odot}$ star and is closer to that of an evolved $\approx$15~$M_{\odot}$ star. However, \citet{boian18} do not rule out an intrinsically fainter progenitor with a dynamic super-Eddington wind \citep[see, e.g.,][]{shaviv01,owocki04,vanmarle09}. 
A well-known example of this is the LBV NGC 2363-V1, %in the giant \ion{H}{2} region NGC\,2366 in the galaxy NGC\,2363, 
which underwent a multiyear, super-Eddington outburst at $\gtrsim$3.5~mag brighter than its comparatively faint progenitor ($M_V \approx -6.5$~mag), though this outburst was significantly hotter ($\gtrsim$11,000~K) than the pre-explosion outbursts of SN\,2015bh (\citealt{drissen01,petit06}, and see \citealt{smith10} for a direct comparison to SN\,2009ip). %\edit1{
$\eta$ Carina also faded below its pre-eruption state in the visible following the ``Great Eruption'' \citep{smith11b}. This may have been associated with dust formation \citep{smith18}, also a ubiquitous feature of merger-related transients \citep[e.g.,][]{martini99,tylenda16,smith16a,blagorodnova17,blagorodnova20}, but again, we do not see evidence for this in SN\,2015bh.%}

%\edit1{
\citet{boian18} describe a possible scenario in which the surviving star settles back into a hot, quiescent LBV state or becomes a Wolf-Rayet star. While our analysis in Section~\ref{sec:SED} does not explicitly rule this out, we do not see clear evidence for a marked change in the SED to support this. This is reflected in the Hertzsprung--Russel diagram (HRD) evolution that we depict in Figure~\ref{fig:cmd_hrd}, in which SN\,2015bh fades well below the luminosity of its massive progenitor. This interpretation is supported by the remarkable similarity with SN\,2009ip, where the optical fading was not accompanied by a UV brightening that would indicate a shift of the SED to higher temperatures \citep{smith22}. SN\,2015bh is also now much fainter than the detection of the believed quiescent progenitor of SN\,2009ip in 1999 ($M_{F606W} \approx -10$\,mag; see Figure~\ref{fig:hst_lcs}). %It is reasonable to suppose that the quiescent progenitor of SN\,2015bh may be similar to that of SN\,2009ip, given the overall similarity of the events themselves. 
%Moreover, the fact that SN\,2015bh fades at a constant $F555W - F814W$ color argues against the scenario where a surviving star is evolving back to a hotter, quiescent state. At the same time, this also rules out a scenario where a luminous surviving star is hidden by late-phase dust formation as the SED should become redder in this case. 
The simplest and, we argue, most likely explanation is that SN\,2015bh was a terminal explosion and the progenitor is now gone.%}

\section{Summary and Conclusions}\label{sec:summary}
We have presented newly analyzed, late-time HST imaging of SN\,2015bh that extends the optical light curves to nearly four years after the explosion. The source fades at a rate of $\approx$0.005~mag~day$^{-1}$ between 2017 and 2018 ($t=596$--975~days) at both $F555W$ and $F814W$ with little to no color evolution. This is strikingly similar to the late-time evolution of SN~2009ip and another member of its class, AT\,2016jbu, and can most likely be attributed to the continued contribution of CSM interaction to the light curves. By 2019 ($t=1404$~days), the $F814W$ light curve slowed its decline rate and may have started to level out, consistent with a scenario where the transient has faded below the level of a persistent, unresolved source at $M_{F814W} \approx -7.0$~mag. The most important result of this work is that the source is now much fainter than its massive, LBV-like progenitor star in broadband optical filters observed with HST in 2008--2009, notably by 3.8~mag at $F814W$, pointing to a scenario where the progenitor did not survive the explosion.
%This strongly indicates that the progenitor did not survive the 2015 explosion or, at least, is now in a dramatically altered state. 

We performed a detailed astrometric analysis showing that the remaining source in 2019 is fully consistent with the position of SN\,2015bh and is highly likely to be associated. One possible scenario is that the observed flux is dominated by an unresolved, scattered-light echo, though the observed color evolution and lack of broad emission features similar to the transient at peak in our 2021 optical spectrum argue against this. %This could be tested with continued, multiband monitoring of the source and detailed modeling of plausible scattering geometries \N{(N: what does "plausible scattering geometries" mean?  do you just mean a dust shell radius, or do you actually mean something about the geometry?)}, in which case any underlying quiescent star or cluster must be even fainter. 
%\edit1{
The observed colors of the remaining source are inconsistent with a main-sequence OB companion star, but we do not rule out the presence of an evolving or inflated companion. Another possible scenario is that the source is settling down to the luminosity level of a relatively small ($\lesssim$3\,pc) and low-mass young cluster. %} 
The late-time spectrum of the site indicates the presence of nearby star-forming regions but shows no evidence of shock-broadened emission from the continued interactions of the SN with dense CSM by $t=2114$~days. Additional multiband photometry (including narrowband H$\alpha$ imaging) with HST or large ground-based facilities will provide context on the progenitor's host population and thereby, valuable constraints on its evolutionary history. %\edit1{
The data presented here disfavor a scenario in which the progenitor survived but is obscured by newly formed dust. A surviving star in transition to a hotter, quiescent state is not yet conclusively ruled out. Still, SN\,2015bh joins SN\,2009ip as an important example of CSM-interacting Type~IIn SNe with increasing evidence for the terminal explosion of their LBV progenitors preceded by decades of eruptive instability. %} 
These objects constitute challenges to models of the evolution of massive stars and continue to provide important constraints in unraveling their tumultuous final years.

%\begin{acknowledgements}
\section{acknowledgements}
We thank the anonymous referee for their comments and suggestions, which helped improve the paper. 

Observations reported here were obtained at the MMT Observatory, a joint facility of the University of Arizona and the Smithsonian Institution.

Based on observations made with the NASA/ESA Hubble Space Telescope, obtained at the Space Telescope Science Institute, which is operated by the Association of Universities for Research in Astronomy, Inc., under NASA contract NAS5-26555. These observations are associated with programs \#HST-SNAP-14668, \#HST-SNAP-15166, and \#HST-GO-15645. Support for program \#HST-GO-15645 was provided by NASA through a grant from the Space Telescope Science Institute, which is operated by the Association of Universities for Research in Astronomy, Inc., under NASA contract NAS5-26555.

Time domain research by the University of Arizona team and D.J.S.\ is supported by NSF grants AST-1821987, 1813466, 1908972, \& 2108032, and by the Heising-Simons Foundation under grant \#2020-1864. J.S.\ acknowledges support from NASA grant HST-GO-15645.003-A and from the Packard Foundation. Research by S.V.\ is supported by NSF grants AST-1813176 and AST-2008108.

Supported by the international Gemini Observatory, a program of NSF's NOIRLab, which is managed by the Association of Universities for Research in Astronomy (AURA) under a cooperative agreement with the National Science Foundation, on behalf of the Gemini partnership of Argentina, Brazil, Canada, Chile, the Republic of Korea, and the United States of America.

\vspace{5mm}
\facilities{HST (ACS, WFC3, WFPC2), MMT (Blue Channel).}

\software{\texttt{AstroDrizzle}, \texttt{TweakReg} (\url{http://drizzlepac.stsci.edu/}; \citealp{hack12}), 
\texttt{DOLPHOT} (\url{http://americano.dolphinsim.com/dolphot/}; \citealp{dolphin00,dolphin16},
IRAF \citep{tody86,tody93},
\texttt{pysynphot} (\url{https://pysynphot.readthedocs.io/en/latest/index.html}; \citealp{pysynphot}),
\texttt{lightcurve\_fitting} (\url{https://griffin-h.github.io/lightcurve_fitting/index.html}; \citealp{hosseinzadeh_lc_zenodo}).}

\appendix

\section{Host Distance and Extinction}\label{appendix:host}
The NASA/IPAC Extragalactic Database (NED)\footnote{NED
is operated by the Jet Propulsion Laboratory, California Institute of Technology,
under contract with the National Aeronautics and Space Administration.} lists several distance estimates to NGC\,2770 based on the Tully--Fisher relation, the most recent of which is listed as $m-M = 32.27 \pm 0.43$~mag ($D = 28.4$\,Mpc; \citealp{sorce14}). \citet{elias-rosa16} adopted a distance modulus of $m-M = 32.33 \pm 0.15$~mag ($D = 29.3$\,Mpc) based on the recession velocity of NGC\,2770 and correcting for the Local Group infall into the Virgo Cluster from NED, while \citet{thone17} assumed $m-M = 32.16$~mag ($D = 27$\,Mpc) and \citet{ofek16} assumed $m-M = 32.38$~mag ($D = 30$\,Mpc). Each of these is consistent with the \citet{sorce14} Tully--Fisher estimate. %and they are all largely consistent with each other, owing to their relatively large uncertainties. 
In this work, we adopt $m-M = 32.3$~mag ($D \approx 28.8$\,Mpc) as an average between the Tully--Fisher and recession-velocity-based estimates. %The primary result of this work, that the SN has faded well below the level of the progenitor star, is independent of the distance assumed.

For the Galactic extinction to the position of SN\,2015bh in NGC\,2770, we adopt the value from the NASA/IPAC Infrared Science Archive (IRSA) of $E(B-V)_{\mathrm{MW}} = 0.02$~mag, based on the \citet{schlafly11} recalibration of the \citet{schlegel98} dust maps. \citet{thone17} estimated the extinction from the host galaxy from the depth of the \ion{Na}{1}~D absorption in a high-resolution spectrum of SN\,2015bh taken on 2015 June 4 and using the relation of \citet{poznanski12} to be $E(B-V)_{\mathrm{host}} = 0.21$~mag. \citet{boian18} obtained an independent and similar estimate of the total (Galactic and host) extinction of $E(B-V) = 0.25$~mag by modeling the 2013 progenitor spectrum. %\N{(N: was this host reddening, or total line of sight including MW? just checking...)}
%JJ: Unclear, but they don't mention MW reddening separately, so it's probably total
In this work, we assume a total extinction to SN\,2015bh of $E(B-V) = 0.23$~mag, consistent with that assumed by \citet{thone17}, and employ the reddening law of \citet{fitzpatrick99} with $R_V = 3.1$ throughout. 

%NED distances 
%
%
%Thone+ 2017: E(B-V)host = 0.21 (Na I D)
%            D = 27 Mpc, mu = 32.16 mag, no reference for it..., typo in mu or DM?
%Elias-Rosa+ 2016: E(B-V)host = 0.21 (Na I D, cite Thone)
%                 D = 29.3 +- 2.1 Mpc, mu = 32.33 +- 0.15 mag 
%                 (NED, recession velocity, local group infall to Virgo)
%Boian & Groh 2018: E(B-V)host = 0.25 (modeling progenitor spectrum)
%                   D (included in uncertainty based on T17 and ER16)
%Ofek+ 2016: D = 30 Mpc mu = 32.38, no reference...
%            ignore host extinction

\section{HST Image Processing and Dolphot Photometry}\label{appendix:phot}
We downloaded the \texttt{calwf3} or \texttt{calacs} calibrated and charge-transfer-efficiency (CTE)-corrected \texttt{flc} frames for the available observations (Section~\ref{sec:phot}) from the Mikulski Archive for Space Telescopes. We processed the images using the \texttt{AstroDrizzle} software package, including automated cosmic-ray rejection, subpixel alignments with \texttt{TweakReg}, and final combination into drizzled mosaics at a pixel scale of $0\farcs05$ for each visit and filter. 
%\J{JS: State here that all mags (or all HST mags) are Vega if that's true.}

We then used \texttt{DOLPHOT} \citep{dolphin00,dolphin16} to obtain PSF-fitting photometry of SN\,2015bh and the sources in its vicinity. We employ the parameter settings used for the Panchromatic Hubble Andromeda Treasury project \citep[PHAT;][]{dalcanton12,williams14}. As inputs to \texttt{DOLPHOT}, we use the CTE-corrected \texttt{flc} frames (preprocessed with \texttt{AstroDrizzle} to flag cosmic-ray hits). %\edit1{
We ran different instrumental setups separately but processed multiple epochs with the same instrument and filter setups together. We used the 2019 drizzled ACS/WFC $F814W$ image as the input reference image for alignment for all runs. %} 
\texttt{DOLPHOT} achieved good alignments at the level of $\approx$0.3\,pixel rms for all of the $F555W$, $F625W$ and $F814W$ images and $\approx$0.4\,pixel rms for the $F438W$ images. \texttt{DOLPHOT} computes and applies aperture corrections to a radius of $0\farcs5$ for the reported photometry. We then applied the appropriate corrections to infinite apertures from \citet{bohlin16} for ACS and from \citet{calamida21} for WFC3.\footnote{See \url{https://www.stsci.edu/hst/instrumentation/wfc3/data-analysis/photometric-calibration}} We find that the statistical uncertainties reported by \texttt{DOLPHOT} are typically much smaller than the spread in individual measurements from each frame that comprises an HST observing visit for a given source. %\edit1{
Therefore, we compute the rms deviations of the individual measurements comprising a given observing visit for all objects ($\approx$500 stars) within a 1000\,pixel radius around SN\,2015bh as a function of magnitude and adopt this as a more realistic estimate of the uncertainty. To check for consistency across the multiple epochs and instrument setups, we also examined the light curves of these stars in our catalogs. We find the 16th--84th percentile deviations between epochs are within $0.1$ and $0.15$\,mag for our $F555W$ and $F814W$ photometry, respectively, indicating good consistency. %}

A clear point source ($\mathtt{objtype} = 1$, indicating a ``good'' star) is detected in all of the 2017 WFC3/UVIS images at the location of SN\,2015bh. This source has faded significantly in 2018 at $F555W$ and $F814W$ (see Figure~\ref{fig:hst_imaging}), confirming it to be the SN. A matching radius of $0.5$ reference-image pixels was used to associate measurements from individual frames in the output source catalogs over the two epochs. A source is also detected by \texttt{DOLPHOT} in our 2019 ACS/WFC $F814W$ catalog at nearly the same location (within $\lesssim 0.2$ pixels) with $\mathrm{S/N} = 14.6$ and $\mathtt{sharpness} = -0.02$, indicating that the object is well fit and consistent with a point-like source (``good'' stars have $-0.3 \leq \mathtt{sharpness} \leq 0.3$). 

\section{Astrometric Analysis}\label{appendix:astrometry}
We performed precise alignments of the 2017 WFC3/UVIS $F555W$ and $F814W$ images, where SN\,2015bh is unambiguously detected, to the deep 2019 ACS/WFC $F814W$ image to confirm that the faint source detected in 2019 is coincident with the SN position. We measured the centroid positions of a set of 42 stars in common between each of the 2017 frames and the 2019 frame and used the IRAF \texttt{geomap} task to determine an alignment solution. We allowed for third-order polynomial fits in both the $x$ and $y$ directions and their cross-terms to account for the field distortion between the sets of images. We then registered the 2017 images to the 2019 images using \texttt{geotran}, and visually examined the resulting images to verify the quality of the registration. We achieved excellent alignments with rms residuals of $0.17$ ACS/WFC pixels (8.5\,mas) in $x$ and $0.19$\,pixels (9.5\,mas) in $y$ for the $F814W$ image and $0.20$ ACS/WFC pixels (10\,mas) in $x$ and $0.26$\,pixels (13\,mas) in $y$ for the $F555W$ image. In Figure~\ref{fig:hst_imaging}, we show the 1$\sigma$ confidence ellipses (assuming 2D Gaussian-distributed alignment residuals) of SN\,2015bh measured in the aligned $F814W$ and $F555W$ 2017 images overlaid on the source detected in 2019. The position of the 2019 source from \texttt{DOLPHOT} (indicated by a white cross in the figure) is fully consistent with the SN position from both 2017 frames within 1$\sigma$.

Based on this analysis, we estimate the probability that the $F814W$ source detected in 2019 is a chance coincidence with an unrelated object. The density of star-like sources within $0\farcs5$ of the position of SN\,2015bh detected by \texttt{DOLPHOT} with $\mathrm{S/N} \geq 5$ is 20.4\,arcsec$^{-2}$. The separation between SN\,2015bh as measured in the aligned 2017 $F814W$ and the 2019 source detected by \texttt{DOLPHOT} is 0.12\,pixels (6\,mas), from which we estimate the likelihood of a chance coincidence with an unrelated source to be only $\approx$0.2\%. For sources as bright as the putative 2019 counterpart, the density drops to 2.5\,arcsec$^{-2}$. The corresponding chance-coincidence probability is then $\approx$0.03\%. Moreover, as the source fades slightly between 2018 and 2019, the likelihood of a chance coincidence with an unrelated, fading variable is even smaller. We therefore conclude that the $F814W$ source detected in 2019 is most likely to be associated with SN\,2015bh. 
%\N{(N: moreover, it is fading... unlikely for a chance coincidence source to also be fading.)}{}

\section{Spectral Analysis of the Nearby \ion{H}{2} Region}\label{appendix:spec}
%\J{JS: If it's more than 280 pc away an association seems very unlikely unless the massive star was ejected at high velocity from a cluster associated with the HII region. I'd say ``likely not directly associated" unless you want to give this more detailed caveat} . 
We fit a Gaussian profile to the H$\alpha$ emission in our spectrum at the location of SN\,2015bh to estimate the integrated line flux at $f_{\mathrm{H\alpha}} = (1.0 \pm 0.1)\times10^{-16}$~erg~s$^{-1}$~cm$^2$ (Galactic extinction correction only). We performed 1000 individual fits, allowing the flux in each spectral bin to vary with Gaussian-distributed noise that encapsulated the 1$\sigma$ fluctuations in line-free regions of the spectrum and the photon-counting source noise in the line itself to estimate the measurement uncertainty. This corresponds to a H$\alpha$ luminosity of $L_{\mathrm{H\alpha}} \approx 1.0$--$1.5\times10^{37}$~erg~s$^{-1}$ depending on the assumed extinction, from no additional host extinction up to the value assumed for SN\,2015bh. For an aperture of the same width but centered on the nearby H$\alpha$ clump, the corresponding luminosities are $L_{\mathrm{H\alpha}} \approx 2.1$--$3.3\times10^{37}$~erg~s$^{-1}$. 

%Maybe this can be moved to an appendix? it's not super important to the rest of the analysis. 
The [\ion{S}{2}]/H$\alpha$ line ratio can also be used to diagnose the presence of emission from an SN remnant. A threshold value [\ion{S}{2}]/H$\alpha > 0.4$ is commonly adopted to indicate the presence of shock heating as the SN remnant interacts with the surrounding medium. Nebular emission from \ion{H}{2} regions typically have [\ion{S}{2}]/H$\alpha \approx 0.1$--0.2 \citep[e.g.,][]{mathewson73,levenson95,long17}, as the photoionization from the UV emission of hot, young stars keeps a high fraction of atomic species in higher-ionization states. We fit Gaussian line profiles to the components of the [\ion{S}{2}] doublet (as described above for H$\alpha$) to measure this ratio in our spectrum at the location of SN\,2015bh. We find [\ion{S}{2}]/H$\alpha = 0.35 \pm 0.1$ for Galactic extinction only or [\ion{S}{2}]/H$\alpha = 0.36 \pm 0.09$ including the host extinction to SN\,2015bh. These values are near the assumed threshold, but they are consistent with other \ion{H}{2} regions on the slit. Thus, we do not find compelling evidence for emission from an SN remnant at the location of SN\,2015bh.

%\J{JS: not sure about the ``emerging" statement here: the standard line ratios are for older, later stage SNRs and aren't relevant for a few-year-old SNR, which will have much broader lines as Nathan says. If there was an SNR spectrum it would be associated with an older SNR, not a 15bh-associated SNR.}
%this is not obviously indicative of a SNR, but is higher than a typical H II reigon. All the sources on the slit in this region have similar ratios or higher. Not sure exactly what to say about this... 
%
% N: this sort of thing is very rough.  people take [S ii] and Halpha images of galaxies and do ratios to identify SNRS.  i think 0.4 is jiust a arbitrary cutoff - SNRs often have higher ratios.  i'm not sure, but the imaging ratios often also have [Nii] in the Halpha filter.  anyway - the key point is that there is not higher velocity emission.  any young SNR would not have such narrow lines.
%\N{(N: the [Sii]/Ha flux ratio is ok, but a key point is that the lines are narrow.  late-time SNe IIn with ongoing interaction and also young SNRs always have shock broadened Halpha lines with widths over 1000 km/s if the interaction is contributing to the emission.)}

\bibliography{jencson}
\bibliographystyle{aasjournal}



\end{document}